\documentclass[letterpaper, 10 pt, journal, twoside]{IEEEtran}
\IEEEoverridecommandlockouts
\usepackage{cite}
\usepackage{amsmath,amssymb,amsfonts}
\usepackage{algorithmic}
\usepackage{graphicx}
\usepackage{wrapfig}
\usepackage{textcomp}
\usepackage{xcolor}
\usepackage{amsmath}

\def\BibTeX{{\rm B\kern-.05em{\sc i\kern-.025em b}\kern-.08em
    T\kern-.1667em\lower.7ex\hbox{E}\kern-.125emX}}
\begin{document}
\pagestyle{empty}

\title{
Pneumatic Units for Logic-based Sequential Excitation (PULSE) in Wearable Haptic Devices\\

\thanks{* Corresponding author: Tania K. Morimoto \\
Jessica Healey, Anoush Sepehri, Michael T. Tolley, and Tania K. Morimoto are with the University of California, San Diego, La Jolla CA 92093, USA}
}

\author{Jessica Healey, Anoush Sepehri, Michael T. Tolley, and Tania K. Morimoto*}


\maketitle
\thispagestyle{empty}

\indent
\begin{abstract}
Soft, wearable robotic devices can deliver haptic feedback to support a wide range of tasks, such as extended reality, training various skills, and rehabilitation. Pneumatic actuation can deliver complex haptic feedback, is lightweight and compliant, and can be incorporated into textiles, making it promising for wearable applications. These soft pneumatic devices, however, typically require a valve and input for each pneumatic actuator, making it challenging to develop fully portable devices for at-home use. In this work we present a pneumatic unit for logic-based sequential excitation (PULSE). The PULSE is a flat, textile-based pneumatic actuator with embedded fluidic logic. By combining these actuators into a fluidic ring oscillator, we decreased the typical amount of required pneumatic inputs for a haptic forearm sleeve by 60\%, with the ability to scale. We built the ring oscillator by optimizing design variables to reach desired periods of oscillation. We demonstrated a set of tactile stroking cues 
with periods ranging from 1.16 to 1.56 s and forces ranging from 1.07 to 2.04 N. We assessed the sleeve’s 
ability to render differentiable, pleasant, and continuous haptic cues in a user study.
The forearm sleeve containing PULSEs successfully delivered four directional cues 
and guided users to target wrist angles with fast reaction times, low overshoot amounts, and a 93.3\% average accuracy of correct initial directions. 
\end{abstract}


\section{Introduction}

Tactile haptic sensations provide feedback through the sense of touch, augmenting a user's experiences for a range of applications, such as teleoperation, extended reality, and prosthetics~\cite{li2026cuthaptics}.
With advancements in sensor and actuator technology, there has been a shift towards wearable form factors to more seamlessly integrate this haptic feedback into people's lives~\cite{fleck2025wearable}. 
While soft haptic devices have the potential to be worn like clothing in daily activities, integrating their actuation and control into portable systems remains a challenge. 

\begin{figure}[t]
\centerline{\includegraphics[width=\columnwidth]{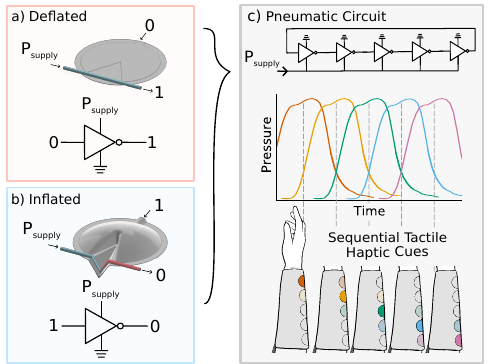}}
\caption{Overview of the wearable haptic sleeve containing PULSEs. The PULSE operates as an inverter and is shown in its (a) deflated and (b) inflated states. (c) With combined logic and actuation in one unit, the PULSE can be easily integrated into a haptic forearm sleeve and deliver sequential tactile cues to users. The top image is a schematic of the PULSE in a pneumatic ring oscillator, the middle image is a representation of the pressure response of the PULSEs, and the bottom image is a representation of the stroking sensation a user will experience, where the intensity of the color shown relates to the intensity of the PULSE's normal force on the skin.}
\label{cover}
\vspace{-15pt}
\end{figure}

Pneumatic actuators are a common choice for wearable haptic systems due to their light weight and compliance compared to electromechanical alternatives, such as motors. To date, wearable haptic devices with pneumatic actuators have been shown to achieve salient directional cues~\cite{raitor2017haptic, agharese2018haptic}. They have also demonstrated the ability to replicate soft human touch via continuous and pleasant stroking gestures~\cite{wu2019haptic}, as well as social gesture cues~\cite{duPasquier2024haptic}. Although these pneumatic wearable devices have shown promise for providing directional or emotional haptic cues to users, they require a valve to control the pneumatic input to each air chamber, resulting in a large electronic backend and limited portability.

Introducing fluidic logic to pneumatic systems can help address the challenge of large amounts of inputs and hardware. One approach is to use the concept of viscous flow to cause a set of air chambers to reach their peak inflation at separate times, based on different amounts of resistance between them.
Vasios et al. have used this concept to tune the inflation rate of a set of bending actuators with one pulsed air input~\cite{vasios2020logic}. Jumet et al. have used a similar fluidic circuit to drive sequential actuation of pneumatic cells in a haptic sleeve \cite{jumet2023logic}. Each cell started to inflate 
at the same time, but reached its peak of inflation at a separate time based on the magnitude of the associated resistor. Determining the direction of the resulting stroking sensation proved somewhat challenging to users. In a stroking sensation, too much delay in sequential actuation can eliminate feelings of continuity, but too little delay can cause the motion to collapse into a single stimulus \cite{henell2024haptic}. We hypothesize that more temporally spaced inflation and deflation profiles will enable clearer perception of the stroking direction and better align with the actuation profiles of previous haptic sleeves that simulated lateral motion with highly rated perceived continuity and pleasantness [5], [10].
    
Pneumatic ring oscillators are a method of fluidic logic that can create more distinct inflation and deflation profiles; they convert a constant input pressure to a time-varying output pressure. Existing designs, however, contain logic separate from their actuation or take 3-dimensional forms not compatible with textile integration. For example, soft ring oscillators with bistable elastomer valves have demonstrated locomotion and have been used for a limb compression device, but the logic elements are separate from the actuation output \cite{preston2019logic, drotman2021logic}. Rajappar et al. presented similar fluidic logic in the form of 2-dimensional ‘textile inverters’---the building blocks of a ring oscillator---that kink tubing to create logic gates \cite{rajappan2022logic}. Although 2-dimensional, these inverters 
still send logic to separate actuators. Some recent devices have combined control and actuation into a single unit. Yang et al. introduced ‘bistable fabric mechanisms’ that produce an oscillatory output with one unit, but their 3-dimensional forms meant for gripping and locomotion limit their integration with textiles \cite{yang2025logic}. A buckling-sheet ring oscillator produced an oscillatory output with units of buckling sheets of acetate that kink tubing \cite{lee2022logic}. While this design combines logic and actuation, its form is not suitable for delivering tactile haptic feedback.

In this work, we present a pneumatic unit for logic-based sequential excitation (PULSE). PULSE is a novel pneumatic actuator designed to be both the logic element in a pneumatic ring oscillator and to provide haptic tactile feedback. The PULSE therefore reduces hardware by combining the logic and actuation required for haptic feedback into one unit. This unit is easy to fabricate, soft, and 2-dimensional when deflated, enabling seamless integration into textiles for wearable use. We create a fluidic ring oscillator with five PULSEs that sequentially inflate and deflate from one constant pneumatic input, and can switch oscillation direction with a second input. This design reduces the number of valves needed for control from five to two (a reduction of 60\%), and can scale to any odd number of PULSEs while still needing only two inputs and valves. Furthermore, we update existing predictive models for the oscillation period and 
use it to optimize physical design parameters, resulting in oscillation periods that provide differentiable haptic cues. We integrate this oscillator into a wearable haptic sleeve to deliver four distinct stroking tactile cues,  while reducing the amount of backend control valves. 
We evaluate user identification and perceived pleasantness and continuity, and demonstrate the ability of the sleeve to guide users through flexion and extension wrist exercises.

\section{Materials and Methods}

\subsection{Actuator Working Principle}


\begin{figure*}[ht]
    \centering
    \includegraphics[width=\textwidth]{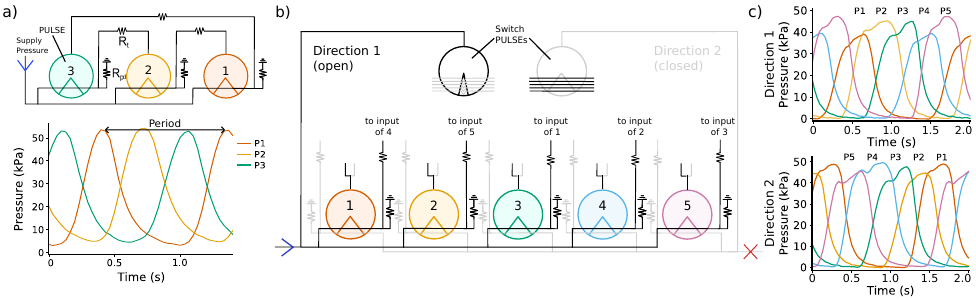}
    \vspace{-20pt}
    \caption{Ring oscillator schematics and their pressure profiles. (a) Schematic of a ring oscillator with three PULSEs and its pressure profile. $R_t$ denotes resistance of the tubing between each actuator and $R_{pd}$ denotes resistance of the pulldown resistor. (b) Schematic of a ring oscillator with five PULSEs and the ability to change direction of oscillation. $R_t$ and $R_{pd}$ are unlabeled but reside in the same places as in (a). The tubes connecting each actuator pass through the `switch' actuators at the top of the image before connecting to the input of the next actuator. (c) The pressure profiles for five PULSEs able to change direction.}
    \label{schematics} 
    \vspace{-15pt}
\end{figure*}

The PULSE is a textile-based pneumatic actuator with a single input and a tube located across the top.
The PULSE creates inverter-like behavior by allowing air to pass through the tube on top when it is in its deflated state (Fig. \ref{cover}a),  and by blocking flow when in its inflated state due to the kinking of the tube (Fig. \ref{cover}b). Importantly, we designed the PULSE such that it can be embedded into fabric layers and produce a normal force on the skin in its inflated state, creating a tactile haptic sensation. 

\subsection{Modeling of Oscillation Period}

When placed in a ring, PULSEs produce the inverter behavior necessary to output sequential inflation with one constant input pressure (Fig.~\ref{schematics}a). Fluidic ring oscillators have been modeled as RC circuits, where resistors represent pneumatic tubing and capacitors represent pneumatic chambers \cite{preston2019logic,lee2022logic}. 
Previous works have used this modeling framework to predict the oscillation period---defined as the time between peaks in an individual actuator's pressure profile---based on input pressure and physical design parameters such as tubing diameter and length. 
Measured parameters, including the buckling pressure and the unbuckling pressure, also contribute to the oscillation period prediction.
The buckling pressure is the pressure at which an actuator blocks off airflow to the following actuator, causing it to deflate. Similarly, the unbuckling pressure is the pressure at which airflow resumes to the following actuator, inflating it. 

To model our system, we made several updates to the existing buckling-sheet ring oscillator model \cite{lee2022logic}. First, we found that the Darcy-Weisbach equation \cite{crowe2009darcy} alone was insufficient for the approximation of the pulldown resistance. Our pulldown resistance, and therefore oscillation period, was largely over-predicted when we increased the length of pulldown resistor tubing, indicating that our pulldown resistance did not have a directly proportional relationship with length. 
The use of connectors to transition from the larger tubing of the ring oscillator to the smaller tubing of the pulldown resistors likely created additional resistance.
To approximate this additional resistance, 
we varied the pulldown resistor's length of tubing, calculated resistance from flow measurements, and took the y-intercept as an additional resistance term to add to the Darcy-Weisbach equation. As a result, our tubing resistances were modeled as:
\begin{equation}
    R_{t} = \frac{128 \mu L_t}{\pi \rho D_t^4}
    \label{tubing_eq}
\end{equation}
\begin{equation}
     R_{pd} = \frac{128 \mu L_{pd}}{\pi \rho D_{pd}^4} + R_{additional}
    \label{tubing_eq_2}
\end{equation}
where $R_{t}$ represents the resistance of the tubing in between each PULSE, $R_{pd}$ represents the resistance of the pulldown resistor, and $R_{additional}$ represents the additional resistance term added to $R_{pd}$ for improved accuracy. $L_t$ and $D_t$ represent the length and inner diameter of the tubing between each PULSE, and $L_{pd}$ and $D_{pd}$ represent the length and inner diameter of the pulldown resistor tubing. 

Similar to the buckling-sheet ring oscillator, our actuators exhibited a continuous shift of pressure, rather than the bistable snap-through transitions found in other soft ring oscillators~\cite{lee2022logic}. 
However, in our system we observed a small increase in unbuckling pressure with increasing supply pressure.
We measured the unbuckling pressure of the PULSE over different input pressures, and performed a linear fit to the experimental data with an $R^2$ value of 0.9572. This linear relationship was then used to predict the unbuckling pressure. The total oscillation period, $t_{period}$, is a combination of the time it takes each PULSE to reach buckling and unbuckling pressures. The time to buckle, $t_{b}$, occurs between a minimum pressure that can be approximated as zero, $P_{min}$, and the buckling pressure, $P_{buckle}$. Time to unbuckle, $t_{u}$, occurs between the effective pressure, $P_{eff}$, and the unbuckling pressure, $P_{unbuckle}$, where the effective pressure accounts for loss from the pulldown resistors \cite{lee2022logic}. We chose the boundary conditions of buckling and unbuckling to reflect this behavior, which leads us to the predictive model represented by: 

\begin{equation}
    t_b = R_{\text{eff}} \times C_{\text{p}} \times \ln \left[ \frac{(P_{\text{min}} - P_{\text{eff}})}{(P_{\text{buckle}} - P_{\text{eff}})} \right]
    \label{buckle_time_eq}
\end{equation}
   
\begin{equation}
    t_{u} = R_{\text{eff}}^* \times C_{\text{p}} \times \ln \left[ \frac{(P_{\text{eff}} - P_{\text{atm}})}{(P_{\text{unbuckle}} - P_{\text{atm}})} \right]
    \label{unbuckle_time_eq}
\end{equation}

\begin{equation}
    t_{\text{period}} = n*(t_b + t_u)
    \label{period_eq}
\end{equation}
where $R_{eff}(R_t,R_{pd},R_p)$ and $R^*_{eff}(R_t,R_{pd})$ are the combined resistances for buckling and unbuckling, respectively~\cite{lee2022logic}. $R_{p}$ and $C_{p}$ are the resistance and capacitance of the PULSE, respectively, $P_{atm}$ is atmospheric pressure, and $n$ is the number of PULSEs. 

Lastly, we fit the lumped parameter resistance and capacitance of the PULSE 
to experimental data as done in previous methods \cite{preston2019logic, lee2022logic}. The $R^2$ value of this fit is 0.9505. The capacitance of the PULSE depends on the size of its air chamber and impacts the period of oscillation of the system. Therefore we characterized the effect of changing the diameter of the PULSE, $D_{p}$, on its capacitance. We measured the step response of PULSEs with different diameters and produced a third-order polynomial fit to relate the PULSE's capacitance to its diameter, with an $R^2$ of 0.9885. 

\subsection{Expanding the Ring Oscillator}

We expanded the current capabilities of soft ring oscillators by sequentially inflating five actuators and switching the direction of the oscillation. Based on preliminary design testing, we selected five PULSEs inflating along the forearm over three PULSEs. Due to the nature of ring oscillators, the system needs an odd number of actuators \cite{preston2019logic}. For five PULSEs to inflate sequentially, we found that each output should be connected to the input of a PULSE three units down the line. The first PULSE's output should be fed into the fourth PULSE and so on (Fig.~\ref{schematics}b). This ensured that the inflation of the first PULSE deflated the fourth PULSE and subsequently inflated the second PULSE next. We observed slight fluctuations in pressure at each pressure peak in this set up (Fig.~\ref{schematics}c). We believe these small variations are due to each actuator sharing the same supply pressure line. Nevertheless, the pressure dip was not large enough to be detected by users.

We also achieved the ability to inflate the same set of PULSEs in the opposite oscillation direction by adding a second path of tubing across them. This tubing path is arranged in the opposite order of the first path, leaving each PULSE with two inputs coming from both directions (Fig.~\ref{schematics}b). With two inputs to each PULSE, air will leak to the atmosphere through the opposite direction unless one path is cut off while the other is in use. Therefore, our design included two `switch' PULSEs, with five tubes laying across them that cut off all paths to the opposite direction while one direction was in use (Fig.~\ref{schematics}b). We can therefore inflate five PULSEs sequentially in both directions with two constant air supplies, where only one supply needs to be on at a time (Fig.~\ref{schematics}c). Since the supply pressure is constant, the typical noise of pneumatic valves switching on and off only occurs when the system changes directions, rather than each time an actuator inflates.

\subsection{Optimization for Desired Haptic Cues}

The period of the ring oscillator is a function of its physical design parameters, which can be tuned before fabrication, and supply pressure, which can be adjusted in real-time. Our goal was to render continuous, pleasant, and effective directional tactile cues with one set of design parameters over a range of supply pressures. We therefore optimized the ring oscillator design to achieve a range of oscillation periods, which enabled a range of frequencies and forces to be rendered to users. 

\begin{figure*}[ht] 
    \centering
    \includegraphics[width=\textwidth]{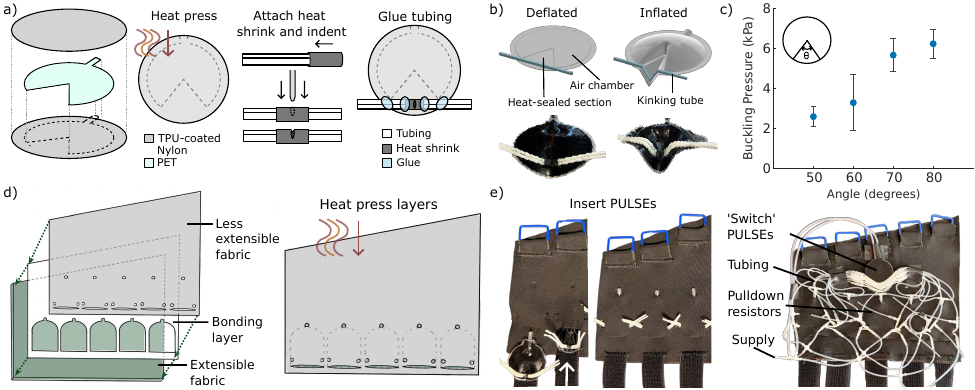}
    \vspace{-15pt}
    \caption{Fabrication of the PULSE and haptic sleeve. (a) To fabricate the PULSE, we heat-pressed the fabric layers together with the PET layer in between. We then sealed heat-shrink around the tubing and indented it to align with the center of the heat-sealed section (where the fold will occur upon inflation). We glued this tubing on top of the fabric layers. (b) Deflated and inflated states of the actuator. (c) The effect of different cutout angles on the PULSE’s buckling pressure. Error bars represent the standard deviations of the buckling pressure from three different measurements. (d) The fabrication process of the haptic sleeve. (e) PULSEs were inserted into the sleeve and tubing was connected externally.}
    \label{fab}  
    \vspace{-15pt}
\end{figure*}

\subsubsection{Desired Haptic Cues}

For affective touch, past works have used haptic forearm sleeves to evaluate the effect of stroking speed on feelings of continuity and pleasantness. These studies resulted in the most pleasant and continuous sensation at a period of 1.2 s or 0.42 s \cite{culbertson2018cue, wu2019haptic}. Literature has shown that stroking motions between 1 and 10~cm/s feel most pleasant on the CT afferents found in hairy skin \cite{tsalamlal2014cue}, however recent studies with pneumatic actuators have found the highest rated pleasantness and continuity to be at 40.3 cm/s or 18 cm/s \cite{duPasquier2024haptic, kommuri2025cue}.
For directional guidance, a fluidically programmed sleeve was able to provide directional stroking cues with a period of 4 s \cite{jumet2023logic}. A study evaluating wrist guidance also noted that pulsing cues provided helpful redundancy and their varying frequencies were more distinguishable than a constant varying strength cue \cite{stanley2012guidance}.
From these results and pilot testing of our system, we predict that a period up to 1.5 s for a stroke length of 16 cm (at least 10.67 cm/s) will provide a clear, pleasant, and continuous stroking sensation.
Lastly, we aimed to render forces between 0.41~N to 6.42~N based on what was previously found to be a comfortable range of pressure stimulation on the forearm~\cite{kodali2023force}, and a distinguishable difference between our maximum and minimum values.


\subsubsection{Design Optimization}

With our predictive model, we used gradient-based optimization to find the optimal physical design parameters, described as $x$ = [$L_t$, $D_t$, $L_{pd}$, $D_{pd}$, $D_p$], for these haptic cues. 
The goal of the optimization problem was to maximize the range of oscillation periods ($r_p$) from our operating range of supply pressures ($P_{supp}$) to create the largest change in haptic sensation. We formulated the optimization as a standard minimization problem by negating the $r_p$ term in the objective function. The supply pressure ranged from 40 kPa, the minimum amount needed for oscillation, to 140 kPa, a value safely under the delamination pressure (172 kPa). The objective function was therefore:
\begin{equation} 
\begin{aligned} 
\min_{x} \quad & -\alpha r_{p}(x, P_{supp}) + (1-\alpha)L_p \\ 
\text{subject to} \quad & 40\,\text{cm} \leq L_t \leq 75\,\text{cm} \\
& 2.0\,\text{cm} \leq L_{pd} \leq 200\,\text{cm} \\
& 0.50\,\text{mm} \leq D_t \leq 0.80\,\text{mm} \\
& 0.50\,\text{mm} \leq D_{pd} \leq 2.0\,\text{mm} \\
& 25\,\text{mm} \leq D_p \leq 40\,\text{mm} \\
& t_{period} \leq 1.5\,\text{s}
\end{aligned} 
\label{opt}
\end{equation} 
where $\alpha$ is the multi-objective function weighting parameter. 
The second term in our objective function aimed to minimize the length of the pulldown resistor to further reduce the bulkiness of the device. We use the gradient-based algorithm Interior Point Optimizer (IPOPT) \cite{wachter2005ipopt} to minimize the objective function and form a Pareto front to select an alpha that provided a reasonable output.

We placed inequality constraints on the PULSE diameters and tubing lengths and diameters to reduce sleeve bulkiness~\eqref{opt}. The lower limit on $L_t$ was the minimum amount of tubing that can connect each PULSE. The pulldown resistors do not pass through the `switch' PULSEs and can be coiled to save space, so they had a broader range of possible values. The constraints on $D_p$ ensured five PULSEs could fit side-by-side along the forearm \cite{gordon2014forearmsize}, and remain at least 40~mm apart center-to-center, which is within the mean values of two-point discrimination for young adults~\cite{nolan1982discrim}.
To create our cues, we placed an upper constraint of 1.5 s on the oscillation period.

\begin{figure*}[ht] 
    \centering
    \includegraphics[width=\textwidth]{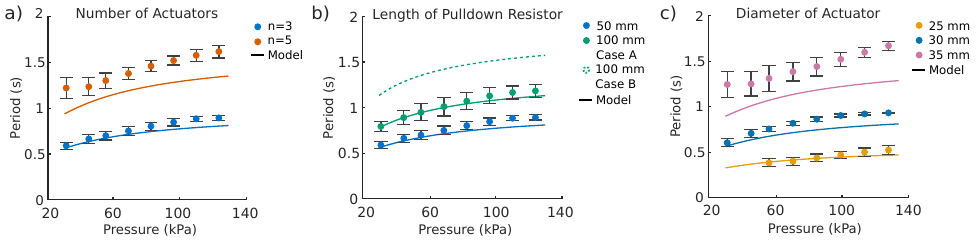}
    \vspace{-15pt}
    \caption{Comparisons between the experimental data of the ring oscillator using the PULSEs and the model. Comparisons shown for (a) the number of actuators, (b) the length of the pulldown resistors, and (c) the diameter of the actuators. Case A in (b) refers to a version of the predictive model without accounting for an additional resistance term in the pulldown resistance, and Case B includes this resistance term. Error bars in (a) and (b) represent the standard deviations of the oscillation period from three different measurements and three different sets of actuators. Error bars in (c) represent the standard deviations of the oscillation period from three different measurements. 
}
    \label{modeling}  
    \vspace{-15pt}
\end{figure*}

\subsection{Actuator Fabrication}

To make the PULSE, we heat-pressed two sheets of TPU-coated Nylon fabric (190 GSM Ripstop Fabric, DIY Packraft) together with a layer of thin PET (7594T13, McMaster-Carr) in between for 120 s at 380$^{\circ}\text{F}$ (Fig.~\ref{fab}a). The shape of the PET layer determines the shape of the air chamber by preventing the fabric layers from sealing together. The result is an open air chamber that can inflate, as well as a heat-sealed section that cannot. The heat-sealed section folds, or buckles, when the air chamber inflates (Fig. \ref{fab}b). The tube glued across this section (5236K501, McMaster-Carr) kinks when the PULSE inflates and unkinks when the PULSE deflates (Fig.~\ref{fab}b).

Two key design features impact the performance of the PULSE: the angle of the heat-sealed section and the reinforcement of the tube's kinking. We formed the PULSE's heat-sealed section with an angled cutout from the PET. When the PULSE inflates, this section folds the fabric and kinks the tubing. We evaluated the buckling pressure of the PULSE across a range of these angles, starting from the smallest angle that could still fit the tubing's glue. Increasing the cutout angle resulted in larger buckling pressures, due to the actuator needing more pressure to fold more fabric while acting on a smaller inflated area. We aimed for a lower buckling pressure so that the system's oscillation could begin at a lower supply pressure. PULSEs with $50^{\circ}$ and $60^{\circ}$ cutouts both had low buckling pressures, so we chose PULSEs with $60^{\circ}$ cutouts to have more room for the glue that secures the tubing (Fig.~\ref{fab}c).

We also reinforced the location where the tubing kinks to enable the PULSE to withstand higher external forces in its inflated, or buckled, state. This reinforcement is beneficial for the PULSE's integration into a haptic sleeve, where the surrounding fabric and donning by a user creates external forces on the PULSE. We found the best method to reinforce this kinking section was to change the cross section of the flexible tubes at the desired kinking location. We accomplished this through sealing heat-shrink over the top of the tubes and creating an indent with a small blunt tip (Fig.~\ref{fab}a). 
The reinforcement enabled the buckled state of the PULSE to withstand higher external forces before unbuckling. 

\subsection{Forearm Sleeve Fabrication}

Our forearm sleeve design combined layers of fabric to create pockets for our actuators and maximize the normal force exerted on the skin. We chose the 
dimensions to align with the minimum average forearm circumference and length so that tightened straps could create a snug fit on most forearms \cite{gordon2014forearmsize}. We found that placing a Velcro strap in line with each actuator produced the best fit on users. To form the sleeve, we laser cut two layers of fabric and one bonding layer into the correct shapes  (Fig.~\ref{fab}d). First, we heat-pressed the bonding layer (HeatnBond Soft Stretch, Therm O Web) onto the bottom layer of fabric (same as inner fabric in \cite{kim2026casaband}, JoAnn Fabrics) for 30 s at 300$^{\circ}\text{F}$. Next, we took the non-stick wrapping off the other side of the bonding layer and heat-pressed the final fabric layer (same as outer fabric in \cite{kim2026casaband}, JoAnn Fabrics) on top for 30 s at 300$^{\circ}\text{F}$. We combined the layers in a pattern that allows the PULSEs to slide into designated pockets with their tubing on the outside of the sleeve (Fig.~\ref{fab}e). Similar to past sleeves that have used variable stiffness in fabrics, the top layer of fabric is less extensible than the bottom layer to allow more downward pressure on the forearm \cite{duPasquier2024haptic, kim2026casaband}. The PULSE weighs 0.91~g and the entire sleeve weighs 50~g.

\section{Results and Discussion}

\subsection{Validation of the Oscillation Period Model}

We demonstrated the agreement between our predictive model and experimental data for key design features. While keeping all other parameters the same, we evaluated the model's performance when changing the number of PULSEs. The RMSE between the experimental data and the model was 0.0611~s [6.70\% error] and 0.229~s [16.0\% error] for three and five PULSEs, respectively (Fig.~\ref{modeling}a). We also demonstrated the effect of changing the pulldown resistor tubing length (Fig.~\ref{modeling}b). The RMSE between the experimental data and the model for 50 mm and 100 mm was 0.0611~s [6.70\% error] and 0.0379~s [2.50\% error], respectively. Case B in Fig.~\ref{modeling}b shows the large over-prediction of oscillation period before adding the additional resistance term to the pulldown resistance (Case A).
Lastly, we tested our model's performance when varying the diameter of the actuators (Fig.~\ref{modeling}c). The RMSE between the experimental data and the model for 25 mm, 30 mm, and 35~mm diameters were 0.0317~s [5.17\% error], 0.105~s [11.9\% error], and 0.301~s [20.8\% error] respectively. Our model had higher errors when increasing the number or diameter of the PULSEs, however the errors are within reasonable variation for pneumatic systems \cite{lee2022logic, ganat2018error}.

\subsection{Optimization of the Oscillation Period}

\begin{figure}[t] 
    \centering
    \includegraphics[width=\columnwidth]{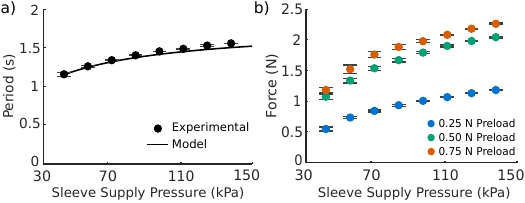}
    \vspace{-15pt}
    \caption{Characterization of the final sleeve. (a) The predicted oscillation period based on the optimization and experimental data from the final sleeve. (b) Force of an individual PULSE within a ring oscillator over supply pressures. The y-axis is the difference between the maximum force measured from the PULSE and the preload force measured from tightening the sleeve. Error bars in both plots represent the standard deviations from three measurements.}
    \label{sleevechar}  
    \vspace{-15pt}
\end{figure}

For the design optimization, we chose a weighting parameter alpha of 0.925, which gave us a balance of a large oscillation period range and a small pulldown resistor length. The length of the tubing between each actuator went to its lower constraint of 40~cm. The inner diameter of the tubing between each actuator went to its upper constraint of 0.80~mm. 
The other optimal parameters were 28~mm for diameter of the PULSE, 0.50~mm for inner diameter of the pulldown resistor, and 5.8~cm for length of the pulldown resistor. These parameters resulted in a predicted period range of 0.370~s, with periods ranging from 1.13~s to 1.50~s. (Fig.~\ref{sleevechar}a).

\subsection{Wearable Haptic Device Characterization}

We characterized the oscillation periods and forces that our final sleeve can render to users. For the sleeve's oscillation periods, we fabricated a ring oscillator with our optimal design parameters and measured its response in the sleeve. With a RMSE between the predicted and experimental values of 0.035~s [2.12\% error] and a range of 0.40~s, its oscillation periods aligned well with the optimized prediction (Fig.~\ref{sleevechar}a). The sleeve could achieve oscillation periods of $1.16 \pm 0.03~\mathrm{s}$ to $1.56 \pm 0.01~\mathrm{s}$ (a stroke speed of 13.8~cm/s to 10.3~cm/s). These stroke speeds are similar to those with highest rated pleasantness and continuity in previous works. Based on preliminary observations, we chose to include a delay between the end and start of the stroking cue for identification. This can be achieved by replacing the pulldown resistor in between the last PULSE and the following PULSE with one of higher resistance. We used 30~cm of tubing to achieve approximately 1~s of delay between the last and first PULSEs. 
If we consider the sensation that the user feels with this delay---time between the first PULSE's peak and the fifth PULSE's peak instead of time between peaks in a single PULSE---we achieve the same range of forces and a range of periods 1.18~s to 1.46~s (or 13.6~cm/s to 11.0~cm/s). These values reach a difference in periods of approximately 0.3~s and remain under 1.5~s. We hypothesized that the difference in oscillation periods and forces from the lowest to highest input pressure would be sufficient for creating two distinct haptic cues in the same oscillation direction. 

To measure force output we placed a single PULSE from the ring oscillator into a one-actuator wide sleeve, and tightened it around a mock wrist with a force sensor. The mock wrist assembly was the same used in \cite{kim2026casaband}, and consisted of a 6-axis force/torque sensor (Mini40, ATI Industrial Automation) attached to a base frame, a force plate, and a skin-like elastomer outer layer (Ecoflex 00-30, Smooth-On). We recorded the maximum force generated by the actuator while the system oscillated. We aimed to tighten the sleeve's straps until they were flush against the user's skin and no tighter, which corresponded to a preload around 0.5~N from tightening the sleeve. We manually tensioned the sleeve to a similar tightness for each user, but in the case of small variations we measured force outputs for preloads of 0.25~N and 0.75~N to show their similarity. In our final sleeve with a tension preload of 0.5~N, PULSEs rendered forces from 1.07~N to 2.04~N over our range of supply pressures (Fig.~\ref{sleevechar}b). This range of forces surpassed the minimum threshold and satisfied the 4\% to 14\% JND for similar forearm indentation tests~\cite{clark2023force}.

\section{User Study}

The goal of this fluidic logic sleeve was to provide users with pleasant and continuous directional cues along the forearm. These tactile sensations can be used for a variety of guidance activities, while leaving the hands free for other tasks. This can be valuable in environments like at-home rehabilitation, where tactile devices have the potential to enhance recovery while enabling more users to experience touch-based assistance without traveling to a clinic \cite{handelzalts2021rehab}. 

To evaluate whether our sleeve's sensations are comfortable and effective,
participants were tasked with identifying a set of directional cues, ranking the sensations' pleasantness and continuity, and reaching a target wrist position through haptic guidance. The study consisted of 10 participants (9 right-handed, 1 left-handed; 3 female, 7 male; age range 23-27) who gave their informed consent, and the protocol was approved by the Institutional Review Board at the University of California, San Diego. 

\begin{figure*}[ht] 
    \centering
    \includegraphics[width=\textwidth]{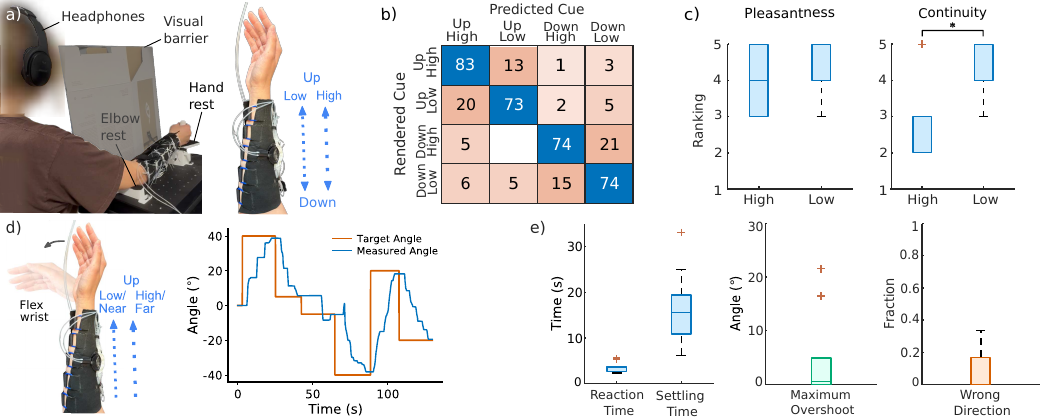}
    \vspace{-15pt}
    \caption{Results for the user study. (a) Image of the test set up and rendered haptic cues, with the participant's vision and hearing blocked and their arm resting on the elbow and wrist supports. (b) Total cue identification results presented as percentages in a confusion matrix. (c) Box plots show the pleasantness and continuity rankings of the High and Low cues on a 5-point Likert scale. (d) Depiction of angle guidance task with representative results from one participant. (e) Box plots show the reaction time, settling time, maximum overshoot, and fraction of wrong initial directions for reaching target angles.}
    \label{userstudy}  
    \vspace{-15pt}
\end{figure*}

The reduced pneumatic inputs for the system allow for simple integration with portable air sources like pumps or compressed air; for this experiment, we used a compressed air line through the wall. The commanded pressure was set by a digital pressure regulator (QB1X, Proportion Air) and a microcontroller (UNO, Arduino). Two solenoid valves (GVP-211C-12D, Automation Direct) controlled the states of the sleeve's two inputs. We 3-D printed an elbow rest and hand rest (Fig. \ref{userstudy}a), which could be fastened down at various distances apart. The hand rest could rotate about the wrist's axis of rotation and contained an encoder (AMT222B-V, Mouser Electronics) to measure the user's position. In the target wrist angle task, an adjustable handle was attached to the hand rest to make flexion and extension movements more comfortable.

\subsection{Cue Identification and Rankings}

In the first part of the study, participants were asked to identify four haptic cues. The two oscillation directions were labeled as Up and Down their arm. The output of the lowest supply pressure---the lowest values of oscillation period and force---were classified as Low intensity, and the output of the highest supply pressure--the highest values of oscillation period and force--were classified as High intensity. Therefore the four haptic cues were Up/High, Up/Low, Down/High, and Down/Low (Fig. \ref{userstudy}a). Participants first underwent a training period of 20 renderings (4 cues, 5 instances each) to learn what each haptic cue felt like. During this training period, the users' hearing and vision were blocked with headphones and a visual barrier, and they were informed of the correct cue if their answer was incorrect (Fig.~\ref{userstudy}a). The cues were pseudorandomized and the user was instructed to provide their best estimate for each rendering
. During the actual study, participants were then tasked with identifying 40 renderings (4 cues, 10 instances each) with no feedback from the operator.

The average identification accuracy of all four haptic cues was 76.0\% (Fig.~\ref{userstudy}b). 
When comparing only direction, participants correctly interpreted 94.5\% of the Up cues and 92.0\% of the Down cues. The most confusion occurred between intensities, where participants correctly interpreted 81.5\% of the High cues and 78.5\% of the Low cues.
Overall, participants distinguished four haptic cues with higher discrimination of direction than intensity. This result suggests that a greater difference between the speeds and forces of the intensities could improve the accuracy of cue identification, which could be achieved through adjusting the optimization weighting parameter and broadening the physical design constraints. While these adjustments would expand the range of speeds and forces, they would also increase the bulkiness of the sleeve.
Our cues were easier to distinguish compared to the previous fluidic logic sleeve that had a 74\% accuracy across all users for distinguishing direction in stroking cues \cite{jumet2023logic}. Notably, this previous system contained one sleeve on each arm for conveying Right vs. Left cues, whereas we convey all four cues along the back of one forearm.

After identification, participants were asked to rank the pleasantness and continuity of the High and Low cues, since Up and Down are the same sensations in different directions. Participants ranked pleasantness and continuity on a 5-point Likert scale from 1 to 5, with 1 being very unpleasant or discrete, 3 being neutral, and 5 being very pleasant or continuous. 
Users found the Low cues to be pleasant (4.3 $\pm$ 0.67) and continuous (4.1 $\pm$ 0.74), whereas the High cues were pleasant (4.1 $\pm$ 0.88) but neutrally continuous (2.8 $\pm$ 0.92) (Fig.~\ref{userstudy}c).
We ran a Wilcoxon Signed Rank Test between the two cues: there was a statistically significant difference between the continuity of the High and Low cues, but no difference between their pleasantness.
This outcome is consistent with previous literature, where trends showed slower stroking speeds had lower perceived continuity but similar perceived pleasantness \cite{culbertson2018cue, wu2019haptic, duPasquier2024haptic}. The rankings indicate that our sleeve provided pleasant and continuous stroking cues to users.

\subsection{Wrist Guidance Task}

Participants were then tasked with reaching a target wrist angle based on the provided haptic cues. The Up cue indicated wrist flexion and the Down cue indicated wrist extension, while the High intensity signaled that they were far from the target angle and the Low intensity signaled that they were close (Fig.~\ref{userstudy}d). The operator aligned the participant's elbow and hand on the rests. Once they reached a deadband of $2.5^{\circ}$ on either side of the target angle, the feedback stopped. After they held the target angle for three seconds, feedback began for a new angle. The participants' hearing and vision were blocked again, and they underwent a training round with a set of six angles ($-40^{\circ}, -20^{\circ}, -5^{\circ}, 5^{\circ}, 20^{\circ}, 40^{\circ}$) in a pseudorandom order. This set was within the average wrist range of motion and had an even amount of flexion and extension targets, with transitions between targets ranging from $5^{\circ}$ to $80^{\circ}$. After this practice round, they completed a formal test with the same set of six angles randomized again.

For wrist guidance, we recorded reaction time, settling time, maximum overshoot, and fraction of wrong initial directions \cite{stanley2012guidance}. Reaction time was defined as the time between the rendering of the cue and the user passing more than one degree from their current position. Settling time was defined as the time between the rendering of the cue and the user reaching the deadband of the target angle, or the length of the total trial minus three seconds. The furthest distance past the target angle was the maximum overshoot (example at $\sim 55$~s in Fig.~\ref{userstudy}d). And lastly, the fraction of wrong initial directions was the proportion of trials in which the user initially moved in the wrong direction (example at $\sim 70$~s in Fig.~\ref{userstudy}d). We computed the average and distribution across all users for these metrics.

Participants had fast average reaction times, 3.58 $\pm$ 1.10~s, and average settling times reasonable for the transition distances, 16.57 $\pm$ 8.06~s (Fig.~\ref{userstudy}e). Participants had low average maximum overshoots at 4.55 $\pm$ $7.88^{\circ}$. Outlier cases demonstrated instances where the participant misinterpreted cues and became lost, with one user reaching an average of $21.6^{\circ}$. Lastly, the fraction of initial wrong directions was 0.0670 $\pm$ 0.117 (Fig.~\ref{userstudy}e). The percentage of total trials with a correct initial direction was 93.3\%. Users often responded quickly and accurately to the rendered cues and were able to correct their wrong directions and overshoots, demonstrating the haptic sleeve's ability to guide users to target angles. 

\section{Conclusion}

Overall, the PULSEs rendered salient tactile haptic cues and create several opportunities for future work. First, while we demonstrated decreased pneumatic valves needed for control, we did not use an on-board air source and leave the technical details of making the system fully portable for future work. Intensity cues were less distinguishable than direction cues, and future work should adjust the optimization's weighting and constraints to prioritize perceptual differences. Future work can also evaluate and improve the robustness of the PULSE to support its prolonged use 
in wearable devices. 
Although we left the tubing outside of the sleeve, this configuration can be refined in future iterations to improve the sleeve's appearance.
Lastly, additional studies could be conducted to evaluate the system's potential for improvements in rehabilitation exercises.

In this work we presented PULSE, a pneumatic unit for logic-based sequential excitation. The PULSE design embeds logic in a haptic tactor with a flat, flexible form factor, reducing backend electronics for soft wearable devices. We characterized its performance within a ring oscillator in a forearm sleeve, presented a predictive model that allowed us to find optimal physical parameters, and evaluated its potential for comfort and guidance through a user study. We are expanding the portability of pneumatic haptic systems by embedding computation directly within soft actuators, and improvements to robustness and textile integration will enable PULSEs to be worn in everyday settings. 

\bibliographystyle{IEEEtran}
\bibliography{biblio}

\end{document}